\documentclass[pra,aps, twocolumn,groupedaddress,showpacs]{revtex4}
\usepackage{amssymb,amsmath,amsfonts,epsfig,graphicx,latexsym,bm,color}

\newcommand{\beq}{\begin{eqnarray}}
\newcommand{\eeq}{\end{eqnarray}}
\newcommand{\bem}{\begin{pmatrix}}
\newcommand{\eem}{\end{pmatrix}}
\newcommand{\nn}{\nonumber}

\newcommand{\f}{\frac}

\newcommand{\dr}[1]{|#1\rangle}

\newcommand{\tb}[1]{\textbf{#1}}
\newcommand{\tr}[1]{\textrm{#1}}

\def\nn{\nonumber}

\def\O{\Omega}

\def\lt{\left}
\def\rt{\right}

\begin{document}
\title{Effective time-independent description of optical lattices with periodic driving}
\author{Andreas Hemmerich}
\affiliation{Institut f\"{u}r Laser-Physik, Universit\"{a}t Hamburg, Luruper Chaussee 149, 22761 Hamburg, Germany}
\date{\today}

\begin{abstract}
For a periodically driven quantum system an effective time-independent Hamiltonian is derived with an eigen-energy spectrum, which in the regime of large driving frequencies  approximates the quasi-energies of the corresponding Floquet Hamiltonian. The effective Hamiltonian is evaluated for the case of optical lattice models in the tight-binding regime subjected to strong periodic driving. Three scenarios are considered: a periodically shifted one-dimensional (1D) lattice, a two-dimensional (2D) square lattice with inversely phased temporal modulation of the well depths of adjacent lattice sites, and a 2D lattice subjected to an array of microscopic rotors commensurate with its plaquette structure. In case of the 1D scenario the rescaling of the tunneling energy, previously considered by Eckardt et al. in Phys. Rev. Lett. 95, 260404 (2005), is reproduced. The 2D lattice with well depth modulation turns out as a generalization of the 1D case. In the 2D case with staggered rotation, the expression previously found in the case of weak driving by Lim et al. in Phys. Rev. Lett. 100, 130402 (2008) is generalized, such that its interpretation in terms of an artificial staggered magnetic field can be extended into the regime of strong driving.\end{abstract}


\maketitle
\date{\today}
\section{Introduction}
Optical lattices are artificial crystalline structures of matter prepared by subjecting ultra-cold neutral atomic gases to spatially periodic light-shift potentials arising in the interference patterns of multiple laser beams \cite{Gry:01}. Quantum degenerate atomic samples arranged in optical lattices allow to study tailored quantum many-body lattice models in a well controlled experimental environment \cite{Lew:08}. While a wealth of lattice geometries are naturally available, a variety of entirely new configurations and tuning options arises if in addition periodic driving is applied. Periodic shaking of a lattice, for example, permits to tune the effective tunneling strength (even to negative values) \cite{Eck:05, Lig:07}, which was recently used to drive a quantum phase transition between a superfluid and a Mott insulator \cite{Zen:09}. Suitably tailored periodic driving schemes allow to implement new building blocks for simulating electronic matter, as for example the effect of the Lorentz-force acting upon the electronic charge in magnetic fields \cite{Sor:05, Lim:08, Lim:10}. This extends the scope of optical lattice models to include intriguing aspects of electronic matter, as for example, quantum Hall physics.

A common approach in the analysis of periodically driven quantum systems is to search for a time-independent effective Hamiltonian with an energy spectrum approximating the quasienergies of the Floquet Hamiltonian of the system \cite{Shi:65, Gro:88, Rah:03}. The accomplishment of this task typically  requires to restrict oneself to specific classes of driving operators. In this article, an effective Hamiltonian is derived for an arbitrary driving operator in the regime of large driving frequencies. This effective Hamiltonian is evaluated for various driven optical lattice models. The lattices are assumed to operate in the tight binding regime described by Hubbard Hamiltonians \cite{Fis:89, Jak:98} with additional external modulation. Firstly, a periodically shifted one-dimensional (1D) lattice is considered. In the regime of strong driving, in accordance with previous work by Eckardt et al. \cite{Eck:05}, a renormalization of the hopping amplitude $J$ is obtained, which permits to tune $J$ even to negative values, a scenario realized in a recent experiment \cite{Lig:07}. In non-bipartite lattice geometries the selective adjustment of negative hopping amplitudes along certain directions of tunneling allows to simulate effects of frustrated magnetism \cite{Eck:10}. Tuning of the energy associated with tunneling is a generic option in driven optical lattice models, not easily realizable in solid state lattices. Secondly, a two-dimensional (2D) square lattice with inversely phased temporal modulation of the well depth of adjacent lattice sites is considered, and a similar rescaling of the hopping amplitude $J$ is found. As a third example, yet unexplored in the regime of strong driving, a 2D square lattice is considered, which is subjected to an array of microscopic rotors commensurate with its plaquette structure. In previous work it was shown that for weak driving this staggered rotation acts to implement the effect of a staggered magnetic field, applying flux with alternating sign to adjacent plaquettes \cite{Lim:08, Lim:10}. Here, it is shown that for strong driving the structure of the effective Hamiltonian and thus its interpretation in terms of a staggered magnetic field is preserved up to non-local tunneling terms, which describe negligible hopping between distant lattice sites. Similarly as in the first example, a rescaling of the tunneling energy arises. Our general expression of the effective Hamiltonian should prove useful for analyzing further cases of interest.

The article is organized as follows: in Sec.~\ref{sec2}, a few relevant elements of Floquet theory are recalled and the connection between the Floquet Hamiltonian of a general periodically driven system and the corresponding time-independent effective Hamiltonian is established. The effective Hamiltonian is expanded into a series of nested commutators. The resulting general expression is applied to the periodically shifted 1D optical lattice in Sec.~\ref{subsec3a}, to the 2D optical lattice with temporal modulation of the well depths in Sec.~\ref{subsec3b}, and to the 2D lattice with staggered rotation in Sec.~\ref{sec4}. Finally, the article is closed with conclusions in Sec.~\ref{conc}. Some straight forward but technical calculations are deferred to the appendix.

\section{Floquet description and effective Hamiltonian}\label{sec2}
According to Floquet's theorem an arbitrary Hamiltonian $H(t)$ with periodic time-dependence ($H(t)=H(t+T)$) operating in some Hilbert space $\mathcal{H}$ possesses a set of $T$-periodic (i.e., $\dr{u_{n}(t)}=\dr{u_{n}(t+T)}$) Floquet states $\dr{u_{n}(t)}$ and a spectrum of quasi-energies $E_n$ determined by the eigenvalue equation for the Floquet Hamiltonian $\mathfrak{H}(t)\equiv H(t) - i \hbar \frac{\partial}{\partial t}$ \cite{Flo:83, Shi:65, Sam:73, Gri:98}
\beq
\label{II.1}
\mathfrak{H}(t) \dr{u_{n}(t)}=E_n \dr{u_{n}(t)}\, .
\eeq
The states $\dr{u_{n}(t)} e^{-iE_{n}t/\hbar}$ form a complete set of solutions to the Schr\"odinger equation $\mathfrak{H}(t) \dr{\psi} = 0$. Recall that  for each $\dr{u_{n}(t)}$ and arbitrary integer $m$ the state $\dr{u_{n}(t),m}æ\equiv \dr{u_{n}(t)} e^{im\Omega t}$ with $\Omega \equiv 2\pi/T$ is itself a Floquet state with quasi-energy $E_{n,m} \equiv E_{n} + m \hbar\Omega$. The solutions to Eq.~(\ref{II.1}) thus display a Brillouin zone-like structure with respect to the time-axis with $E_{n}$ to be chosen within the first zone $[-\hbar\Omega/2,\hbar\Omega/2]$. The states $\dr{u_{n}(t),m}$  with $E_{n} \in [-\hbar\Omega/2,\hbar\Omega/2]$ form an orthonormal basis in the composite Hilbert space $\mathcal{H} \otimes \mathcal{H}_T $,  where $\mathcal{H}_T$ is the space of $T$-periodic complex-valued functions. Hence, $\delta_{n,n'} \delta_{m,m'} =ææ\langle \langle u_{n}(t),m | u_{n'}(t),m' \rangle \rangle_{T}$ with $\langle \langle \phi(t) | \psi(t) \rangle \rangle_{T} \equiv \frac{1}{T}\int_{0}^{T} dt \langle \phi(t) | \psi(t) \rangle \rangle_{T}$ denoting the scalar product in $\mathcal{H} \otimes \mathcal{H}_T $. Equation~(\ref{II.1}) may thus be considered as an eigenvalue problem in the composite Hilbert space $\mathcal{H} \otimes \mathcal{H}_T $ \cite{Sam:73}.

For an arbitrary stationary orthonormal basis $\dr{n}$ of $\mathcal{H}$ and an arbitrary time-periodic Hermitian operator $F=F^{\dagger}$ with $F(t) = F(t+T)$ one may define an orthonormal basis of the composite Hilbert space $\mathcal{H} \otimes \mathcal{H}_T$ by $\dr{n(t),m} = U_{F,m}(t) \dr{n}$ with $U_{F,m}(t) \equiv e^{-iF(t) + i m \Omega t}$ and arbitrary integer $m$.
Defining 
\beq
\label{II.2}
\mathfrak{H}_{F}^{(m,m')}&\equiv& U_{F,m}^{\dagger}(t) \mathfrak{H}(t) U_{F,m'}(t) \\ \nn \\
H_{\textrm{eff}}æ&\equiv& \langle \mathfrak{H}_{F}^{(0,0)} \rangle_{T} \, ,
\eeq
where $\langle \dots \rangle_{T}$ denotes time averaging over the periode $T$, it is straight forward to verify
\beq
\label{II.3}
\langle \langle n(t),m | \mathfrak{H}(t) | n'(t),m' \rangle \rangle_{T}= \qquad\qquad\qquad \nn \\  \\ \nn
\delta_{m,m'} \left( \langle n | H_{\textrm{eff}} | n' \rangle + m \hbar \Omega \right) \qquad  \\ \nn
+ \left( 1-\delta_{m,m'}\right) \langle n | \langle e^{i(m'-m)\Omega t}\mathfrak{H}_{F}^{(0,0)} \rangle_{T} | n' \rangle \, .
\eeq
According to Eq.~(\ref{II.3}) the matrix elements of $\mathfrak{H}(t)$ in the basis $\dr{n(t),m}$ of the composite space $\mathcal{H} \otimes \mathcal{H}_T$ display a block structure with diagonal $(m=m')$  blocks $\langle n | H_{\textrm{eff}} | n' \rangle + m \hbar \Omega$ energetically separated by multiples of $\hbar \Omega$ and off-diagonal $(m \neq m')$ blocks $\langle n | \langle e^{i(m'-m)\Omega t}\mathfrak{H}_{F}^{(0,0)} \rangle_{T} | n' \rangle$ coupling different diagonal blocks. Following a well-known result of perturbation theory, if
\beq
\label{II.4}
||\langle e^{i(m'-m)\Omega t} \mathfrak{H}_{F}^{(0,0)}\rangle_T|| \ll \hbar \Omega  
\eeq
is satisfied for all $m\neq m'$ with $||A|| \equiv \textrm{Max}\{ |\langle n| A|m \rangle|: n,m \}$, the off-diagonal couplings can be neglected yielding the approximation 
\beq
\label{II.5}
\langle \langle n(t),m | \mathfrak{H}(t) | n'(t),m' \rangle \rangle_{T} \approx \qquad\qquad\qquad  \\ \nn
\delta_{m,m'} \left( \langle n | H_{\textrm{eff}} | n' \rangle + m \hbar \Omega \right) \, . \qquad 
\eeq 
Thus, within the range of validity of condition (\ref{II.4}), and given that the energy levels resulting from blocks with different $m$ do not mix, i.e., if
\beq
\label{II.6}
||H_{\textrm{eff}}|| \ll \hbar \Omega \,  
\eeq
holds, equation~(\ref{II.5}) shows that the quasi-energy spectrum of $H(t)$ within the first Brillouin zone $(m=m'=0)$ coincides with the energy spectrum of the time-averaged effective Hamiltonian $H_{\textrm{eff}}$. Consequently, within the sub-space associated to the first energy band $E_{n,m=0}$ the time-dependent Hamiltonian $H(t)$ and the time-independent Hamiltonian $H_{\textrm{eff}}$ are equivalent. 

The crucial task in practical applications is to identify a suitable operator $F$ compatible with the constraints imposed by conditions (\ref{II.4}) and (\ref{II.6}). A useful recipe in this respect is to decompose the Hamiltonian $H(t) = H_{<}(t) + H_{>}(t)$ into a weakly driven part $H_{<}(t)$, satisfying $||H_{<}(t)|| \ll \hbar \Omega$, and a strongly driven part $H_{>}(t)$, and then to choose $F$ in order to integrate out only $H_{>}(t)$, i.e., $\hbar F(t) \equiv \int_{0}^{t}ds\,H_{>}(s)$. A useful expansion of $\mathfrak{H}_{F}^{(0,0)}$ in terms of multiple commutators involving $F$ and $F'\equiv \partial F/\partial t$ can be derived. Defining the multiple commutator between operators $A$ and $B$ of order $n+1$ by the recursion $\left[A,B\right]_{n+1} \equiv \left[A,\left[A,B\right]_{n}\right]$ and $\left[A,B\right]_{0} \equiv B$, the practical relations 
\beq
\label{II.7}
e^{iF}\frac{\partial}{\partial t} e^{-iF}æ&=& -æ\sum_{n=0}^{\infty} \frac{i^{n+1}}{(n+1)!}\left[F,F'\right]_{n}
\nn \\  \\  \nn
e^{iF}G\,e^{-iF} &=& \sum_{n=0}^{\infty} \frac{i^n}{n!}\left[F,G\right]_{n}
\eeq
hold for arbitrary operators $F$ and $G$ \cite{Eng:63, Czi:92}. Inserting Eq.~(\ref{II.7}) with $G\equiv H(t)$ into Eq.~(\ref{II.2}) for $m=m'=0$ yields 
\beq
\label{II.8}
\mathfrak{H}_{F}^{(0,0)} = \sum_{n=0}^{\infty} \frac{i^n}{n!} \left(\left[F(t),H(t) \right]_{n} - \frac{\hbar}{n+1}\left[F(t), F'(t)\right]_{n}\right) \nn \\   
\eeq

Henceforth, the decomposition $H(t) = H_{<}(t) + H_{>}(t)$ is applied and $F(t)$ is chosen to satisfy $\hbar F'(t) = H_{>}(t)$ One may then rewrite Eq.~(\ref{II.8}) as
\beq
\label{II.9}
\mathfrak{H}_{F}^{(0,0)} = \qquad\qquad\qquad\qquad\qquad\qquad\qquad\qquad \\
\sum_{n=0}^{\infty} \frac{i^n}{n!} \left(\left[F(t), H_{<}(t) \right]_{n} + \frac{n}{n+1}\left[F(t), H_{>}(t)\right]_{n}\right) \nn
\eeq
In the following sections expression (\ref{II.9}) will be applied to several examples of driven optical lattices and conditions (\ref{II.4}) and (\ref{II.6}) are inspected to determine the range of validity for approximating $H(t)$ by $H_{\textrm{eff}}$. 

\section{Dynamical control of tunneling in optical lattices}\label{sec3}
\subsection{Periodic shaking of 1D optical lattice}\label{subsec3a}
First, the periodical shaking of a 1D optical lattice is considered, which allows to suppress tunneling and even simulate negative tunneling energies. This scenario has been previously investigated theoretically in Ref.~\cite{Eck:05} and experimentally in Ref.~\cite{Lig:07} and thus permits a useful test of Eq.~(\ref{II.9}). The Hamiltonian is written as $H(t) = H_0 + W(t)$. Within the tight binding regime the time-independent Hamiltonian is the 1D Bose-Hubbard Hamiltonian $H_0 = -J\,T_{+}  +  H_{\textrm{int}}$ with the tunneling operators $T_{\pm} \equiv \sum_{\langle  \nu,\mu \rangle} c_{ \nu}^{\dag}c_{\mu} \pm c_{\mu}^{\dag}c_{\nu}$ and the onsite interaction $H_{\textrm{int}} = \frac{U}{2}\sum_{ \nu} \hat{n}_{ \nu} (\hat{n}_{\nu} -1)$ \cite{Fis:89,Jak:98}. The periodic driving operator reads $W(t) = 2 Q\, \cos(\Omega t)$ with $Q \equiv \chi \sum_{\nu} \nu \hat{n}_{\nu}$. Here, $c_{\nu}$ denotes the bosonic anihilation operator at site $\nu$, $\hat{n}_{\nu}$ is the corresponding particle number operator, and $\langle  \nu,\mu \rangle$ indicates summation over pairs of nearest neighbor sites. The parameters $J$, $U$, and $\chi$ quantify the tunneling strength, the on-site repulsion energy per particle, and the modulation strength, respectively. The operator $Q$ acts to introduce a constant gradient of the chemical potential and thus a constant force $\chi/d$, where $d$ is the lattice constant. Hence, the driving term $W(t)$ represents a tilt of the lattice with harmonic time-dependence. Experimentally, $W(t)$ is realized by periodically shifting the 1D standing wave forming the optical potential and transforming to the co-moving frame of reference \cite{Mad:98}. 

It has been shown recently in Ref.~\cite{Eck:05} that for sufficiently high driving frequencies the driven system $H(t)$ behaves similarly as the undriven system $H_0$, but with the tunneling matrix element $J$ replaced by the effective matrix element $J J_0(2 \chi / \hbar \Omega)$, where $J_0$ denotes the Bessel function of order zero. Notably, since $J_0$ can take negative values, negative values of the effective tunneling strength should become possible, a prediction confirmed experimentally \cite{Lig:07}. This result is readily reproduced by means of Eq.~(\ref{II.9}). Choosing $F(t)= 2 Q\, \sin(\Omega t)/(\hbar \Omega)$ yields $\hbar F'(t) = W(t)$ and $\left[F(t), W(t)\right]=0$. Eq.~(\ref{II.9}) with $H_{<}(t) \equiv H_0$ and $H_{>}(t) \equiv W(t)$ thus simplifies to
\beq
\label{IIIA.1}
\mathfrak{H}_{F}^{(0,0)} = \, \sum_{n=0}^{\infty} \frac{1}{n!} \left(\frac{2 i \sin(\Omega t)}{\hbar \Omega}\right)^{n}   \left[Q, H_{0}\right]_{n} \, . 
\\  \nn
\eeq
Use of $\left[Q, H_{\textrm{int}}\right]=0$ and evaluation of the commutators $\left[Q,T_{\pm}\right] = \chi \,T_{\mp}$ and thus $\left[Q,T_{\pm}\right]_{n} = \chi^{n} \,T_{(-1)^n \pm }$ leads to
\beq
\label{IIIA.2}
\mathfrak{H}_{F}^{(0,0)} = \,H_{\textrm{int}}
&-&J \, \sum_{n=0}^{\infty} \frac{1}{(2n)!} \left(\frac{2 i \chi \sin(\Omega t)}{\hbar \Omega}\right)^{2n}  T_{+}\,
\\  \nn
&-&J \, \sum_{n=0}^{\infty} \frac{1}{(2n+1)!} \left(\frac{2 i \chi \sin(\Omega t)}{\hbar \Omega}\right)^{2n+1}  T_{-}\, .
\eeq
Applying the relations
\beq
\label{IIIA.3}
\langle e^{i 2m z}\sin^{2n}(z)\rangle &=& \frac{(-1)^m (2n)!}{2^{2n} (n+m)! (n-m)!}  \\
\langle e^{i(2m+1)z}\sin^{2n+1}(z)\rangle &=& i \frac{(-1)^m (2n+1)!}{2^{2n+1} (n+m+1)! (n-m)!} \nn \\
\langle e^{i 2m z}\sin^{2n+1}(z)\rangle &=&\langle e^{i(2m+1)z}\sin^{2n}(z)\rangle = 0 \, , \nn 
\eeq
for integers $n,m \leq 0$ and $(n-m)! \equiv 0$ if $n < m$ and $\langle \dots \rangle$ denoting the average over $2\pi$ with respect to $z$,
and making use of the power expansion of the $m$-th order Bessel function
\beq
\label{IIIA.4}
J_m(z)=\sum_{n=0}^\infty  \frac{(-1)^n z^{2n+m}}{2^{2n+m}n!(m+n)!} 
\eeq
finally yields
\beq
\label{IIIA.5}
\langle e^{i2(m'-m)\Omega t} \mathfrak{H}_{F}^{(0,0)}\rangle_T  &=& -J \, J_{2(m'-m)}\left(\frac{2\chi}{\hbar \Omega}\right) \, T_{+}  \nn  \\ 
&+&  \, \, \delta_{m',m} \, H_{\textrm{int}} \, , \nn  \\ \\ \nn \\ \nn 
\langle e^{i(2(m'-m)+1)\Omega t} \mathfrak{H}_{F}^{(0,0)}\rangle_T  &=& J \, J_{2(m'-m)+1}\left(\frac{2\chi}{\hbar \Omega}\right) \, T_{-} \, , \nn\eeq
and (setting $m=m'$)
\beq
\label{IIIA.6}
H_{\textrm{eff}}æ=  -J \, J_{0}\left(\frac{2\chi}{\hbar \Omega}\right) \, T_{+}  \, \, +  \, \,  H_{\textrm{int}} \, . 
\eeq
Equation~(\ref{IIIA.6}) confirms the rescaling of the tunneling energy derived in Ref.~\cite{Eck:05}. Finally, the range of applicability of $H_{\textrm{eff}}$ remains to be discussed via inspection of Eqs.~(\ref{II.4}) and ~(\ref{II.6}). For increasing number of particles in the optical lattice, these conditions avoiding any coupling between different Floquet bands become increasingly hard to satisfy, because multiple excitations (e.g., multiple particle-hole excitations due to collisions) then may bridge increasing energy intervals. However, as has been pointed out in Ref.\cite{Eck:08}, such higher order excitations yield only small couplings acting on time-scales too long to be relevant in typical experiments. Thus, it suffices to require that the single particle energy scaling factors $|J \, J_{m}\left(2\chi/\hbar \Omega\right)|$ and $U$ in Eqs.~(\ref{IIIA.5}) and Eqs.~(\ref{IIIA.6}) do not exceed $\hbar \Omega$. Because $J_{m}(z) \leq 1$ for arbitrary integers $m$, a sufficient condition for any value of the driving strength $\chi$ is $J \ll \hbar \Omega$ and $U \ll \hbar \Omega$.

In the derivation of Eq.~(\ref{IIIA.6}) a harmonic modulation $W(t) \propto \cos(\Omega t)$ has been assumed, however, other time-dependences yield a similar renormalization of the hopping strength $J$. If, for example, $\cos(\Omega t)$ is replaced by a rectangular modulation function $f(t) \equiv (-1)^{\nu}$ for $-\pi/2 + \nu \pi \leq \omega t < \pi/2 + \nu \pi$, where $\nu$ runs through all integers, the zero order Bessel-function $J_{0}$ in Eq.~(\ref{IIIA.6}) is to be replaced by the sinc function.

\subsection{Well depth modulation in 2D optical lattice}\label{subsec3b}
\begin{figure}
\includegraphics[scale = .35, angle=0, origin=c]{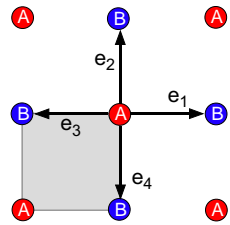}
\caption{\label{Fig.1} (color online) 
Decomposition of 2D square lattice into two sublattices indicated by $\mathbb{A}$ (red) and $\mathbb{B}$ (blue). The vectors $\tb{e}_{\nu}$, $\nu=1,2,3,4$, connect an $\mathbb{A}$-site to its four nearest neighboring $\mathbb{B}$-sites. The grey area denotes a $\lambda/2 \times \lambda/2$-sized plaquette.}
\end{figure}

Next, a 2D square optical lattice is considered, composed of two sublattices as illustrated in Fig.~\ref{Fig.1}. The difference between the well depths of the $\mathbb{A}$ and the $\mathbb{B}$ sites is assumed to be modulated harmonically. Experimentally, this scenario arises if two standing light waves with wavelength $\lambda$ and parallely oriented linear polarizations are crossed by superimposing the two branches of a Michelson interferometer. Control of the optical path length difference of the interferometer allows to adjust the difference $\theta$ between the time-phases of the two standing waves. The resulting optical potential is $V(x,y) = -\bar V_0 \, |e^{i \theta/2}\sin(kx) +e^{-i \theta/2}\sin(ky)|^2$ \cite{Hem:92} with $k=2\pi/\lambda$. It is straight forward to implement the temporal modulation $\theta =  (1+ \kappa \cos(\Omega t))\, \pi/2$. For $\kappa \ll 1$ this yields an overall potential $V(x,y,t)  =  -\bar V_0 \left(\sin^2(kx) +\sin^2(ky)\right) -\bar V_0 \pi \sin(kx)\sin(ky) \cos(\Omega t)$ composed of a stationary 2D square lattice with mean well depth $\bar V_0$ and the desired harmonic modulation of the well depth difference of $\mathbb{A}$ and $\mathbb{B}$ sites. Within a tight binding description in terms of a driven Hubbard model a calculation similar to those in Ref.~\cite{Jak:98, Hem:07, Lim:10} yields the Hamiltonian
\beq
\label{IIIB.1}
H(t) &=& -J T_{+} + H_{\tr{int}} + W(t)\,, \nn \\ \nn \\
T_{\pm} &\equiv& \sum_{\tb{r}\in \mathbb{A} }\sum_{\nu\in {1,2,3,4}}a_{\tb{r}} b_{\tb{r}+\tb{e}_{\nu}}^{\dag} \pm a_{\tb{r}}^{\dag} b_{\tb{r}+\tb{e}_{\nu}}\,, \nn \\ \nn \\ 
H_{\tr{int}}&=& \f{U}{2}\sum_{\tb{r}\in \mathbb{A} \oplus \mathbb{B}}\hat{n}_{\tb{r}}\,(\hat{n}_{\tb{r}}-1)\,,  \\  \nn \\
W(t) &=& 2 Q \cos{\O t}, \,\, Q \equiv  -\frac{\chi}{2} \sum_{\tb{r}\in \mathbb{A}} (\hat{n}_{\tb{r}}-\hat{n}_{\tb{r}+\tb{e}_1})\,\,.\nn
\eeq
Here, $a_{\tb{r}}$ and  $b_{\tb{r}+\tb{e}_{\nu}}$ denote the bosonic anihilation operators at sites $\tb{r}$ and $\tb{r}+\tb{e}_{\nu}$ of sublattices $\mathbb{A}$ and $\mathbb{B}$, respectively. The corresponding particle number operators are $\hat{n}_{\tb{r}}  \equiv a^\dag_{\tb{r}}a_{\tb{r}}$ and $\hat{n}_{\tb{r}+\tb{e}_1} \equiv b^\dag_{\tb{r}+\tb{e}_1} b_{\tb{r}+\tb{e}_1}$. The four vectors $\tb{e}_{\nu}$, $\nu=1,2,3,4$, connect an $\mathbb{A}$-site to its four nearest neighboring $\mathbb{B}$-sites, according to Fig.~\ref{Fig.1}. In analogy to the previous section, the parameters $J$ and $U$ quantify, respectively, the tunneling strength and the on-site repulsion energy per particle of the conventional Bose-Hubbard Hamiltonian for a 2D square lattice. The modulation strength is $\chiæ=æ- \pi \kappa \bar V_0 \int dx dy \, |w(x,y)|^2 \cos(k x)\cos(k y)$, where $w(x,y)$ denotes the Wannier function of the lowest band. Although the definitions here refer to a 2D lattice, a close formal analogy to the previously discussed 1D case can be observed. In fact, with $F= 2 Q\, \sin(\Omega t)/(\hbar \Omega)$ and the definitions of Eq.~(\ref{IIIB.1}) the same commutation relations as those found in subsection~\ref{subsec3a} are recovered, i.e., $\left[F(t), W(t)\right]=0$, $\left[Q, H_{\textrm{int}}\right]=0$ and $\left[Q,T_{\pm}\right]_{n} = \chi^{n} \,T_{(-1)^n \pm}$. Consequently, equations (\ref{IIIA.2}), (\ref{IIIA.5}) and (\ref{IIIA.6}) also hold for the 2D scenario of Eq.~(\ref{IIIB.1}). As in the 1D case, one may thus use the modulation parameter $\chi$ in order to suppress tunneling or even adjust negative values of the tunneling strength. Arguments analogue to those used at the end of subsection~\ref{subsec3a} show that  for any value of the driving strength $\chi$ the time-independent effective description is justified if $J,U \ll \hbar \Omega$.

\section{2D optical lattice with staggered rotation}\label{sec4}
\begin{figure}
\includegraphics[scale = 0.5, angle=0, origin=c]{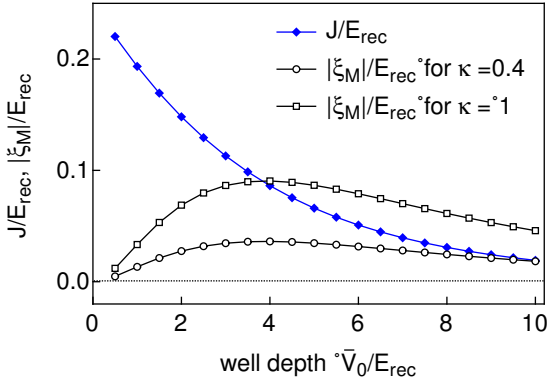}
\caption{\label{Fig.2} (color online) Plot of $J$ and $|\xi_M|$ versus the lattice well depth $\bar V_0$ for $\kappa = 0.4$ and $\kappa = 1$. All quantities are scaled to the recoil energy $E_{\textrm{rec}} \equiv \hbar^2 k^2/2m$ of the photons providing the optical lattice potential. $4J$ is approximated by the energy width of the lowest band.} 
\end{figure}

In this section the square optical lattice in Eq.~(\ref{IIIB.1}) of Sec.~\ref{subsec3b} is considered, however with a geometry of the driving term, specifically designed to apply angular momentum with alternating sign to neighboring plaquettes. It has been shown in Refs.~\cite{Hem:07, Oel:09} that this can be experimentally realized in a bichromatic optical lattice, produced in an optical set-up comprising two nested Michelson interferometers. This yields an optical potential $V(x,y,t) = V_{\tr{L}}(x,y) + V_{\tr{R}}(x,y,t)$ consisting of the stationary square lattice $V_{\tr{L}}(x,y) = -\bar V_0 \left(\sin^2(kx) +\sin^2(ky)\right)$ of Eq.~(\ref{IIIB.1}) and a temporal modulation $V_{\tr{R}}(x,y,t) = \kappa V_{\tr{L}}(x,y) \cos(2 S(x,y)-\O t)$ with $\kappa$ adjustable within $[0,1]$ and
\beq
\label{IV.1}
S(x,y) \equiv \tan^{-1}\biggl\{\f{\sin(k x)-\sin(k y)}{\sin(k x)+\sin(k y)}\biggr\},
\eeq
which acts as an array of microscopic rotors, each centered in an individual plaquette. The well depth $\bar V_0$ scales linearly with the overall intensity of the lattice beams while $\kappa$ is adjusted via the intensity ratio of the two frequency componets of the bichromatic lattice. A description in terms of a driven Hubbard model leads to (cf. Ref.~\cite{Hem:07})
\beq
\label{IV.2}
H(t) &=& -J T_{+}  + H_{\tr{int}}  + W(t)\,, \nn \\ \nn \\
W(t) &=& \xi_N N \cos(\O t) -\xi_M M \sin(\O t)\,, \nn \\ \nn \\
N&\equiv& \sum_{\tb{r}\in \mathbb{A}} (\hat{n}_{\tb{r}}-\hat{n}_{\tb{r}+\tb{e}_1})\,,   \\
M&\equiv&\sum_{\tb{r}\in \mathbb{A}, \nu=1-4}(-1)^{\nu+1} (a^\dag_{\tb{r}}b_{\tb{r}+e_\nu}+ \tr{H.c.})\, . \nn
\eeq
The operators $T_{\pm}$ and $H_{\tr{int}}$ are the same as in Eq.~(\ref{IIIB.1}). The modulation strength parameters are given by $\xi_N = 2 \kappa \bar V_0 \int dx dy \, |w(x,y)|^2 \cos(k x)\cos(k y)$ and $\xi_M= \kappa \bar V_0 \int dx dy \, w^*(x+\lambda/4,y)[\sin^2(k x)-\cos^2(k y)]w(x-\lambda/4,y)$. Since the 2D Wannier function of the lowest band of the square lattice $V_{\tr{L}}(x,y)$ factorizes ($w(x,y)=w(x)w(y)$), one obtains the simplified expressions $\xi_N = 2 \kappa \bar V_0 \left[ \int dx \, |w(x)|^2 \cos(k x) \right]^2$ and $\xi_M= \kappa \bar V_0 \int dx \, w^*(x+\lambda/4) \sin^2(k x) w(x-\lambda/4)$. Because $w(x)$ is peaked around $x=0$, $\xi_N$ is on the order of $2 \kappa \bar V_0$. In contrast, $\xi_M$ exhibits the same order of magnitude as $J$ because it involves a tunneling integral, which receives its main contributions from the small side lobes of the Wannier function. This is confirmed by a band structure calculation including the first five bands as is shown in Fig.~\ref{Fig.2}, where $J$ and $|\xi_M|$ are plotted versus the lattice well depth $\bar V_0$ for $\kappa = 0.4$ and its maximum possible value $\kappa = 1$. For $\kappa >\approx 0.4$ the value of $\xi_M$ can exceed that of $J$ for increasing $\bar V_0$, which would lead to negative overall hopping amplitudes of $H(t)$ for certain fractions of the modulation cycle. This indicates that the description in terms of the lowest band Wannier function of the stationary lattice $V_L$ becomes questionable then.

\begin{figure}
\includegraphics[scale = 0.35, angle=0, origin=c]{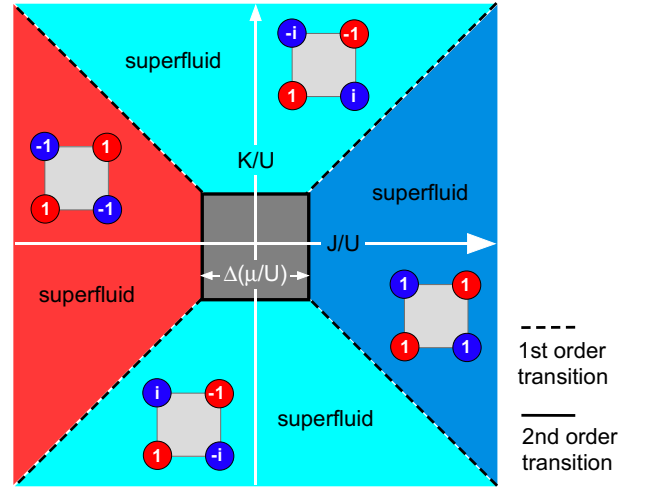}
\caption{\label{Fig.3} (color online) 
Zero temperature phase diagram of the Hamiltonian $-J T_{+} - i K \frac{1}{2} \left[M, N \right] + H_{\textrm{int}}$ within the ($J$,$K$)-plane calculated according to Ref.~\cite{Lim:08} for fixed chemical potential $\mu / U$. Within a rectangular region (grey) around the origin of the ($J$,$K$)-plane a Mott insulator is formed. It is surrounded by four superfluid phases, each with an order parameter characterized by different values at the four lattice sites at the corners of a plaquette as indicated by the small associated pictograms. The width of the Mott region is $\Delta(\mu/U)$, where $\Delta(z) \equiv [z(2g-1-z) + g(g-1)]/ [4(1+z)]$ for a filling fraction $g \in \{1,2,...\}$ and $z \in [g-1,g]$. Solid (dashed) black lines denote second (first) order phase boundaries.}
\end{figure}

For weak driving  ($\xi_M  \ll \hbar \Omega$ and $\xi_N  \ll \hbar \Omega$) a corresponding time-averaged effective Hamiltonian $H_{\textrm{eff}} = -J T_{+} - i K \frac{1}{2} \left[M, N \right]$ with $K \equiv \xi_N \xi_M/(\hbar \Omega)$ has been introduced in Ref.~\cite{Lim:08}. The interest in this scenario arises because this Hamiltonian mimics the action of a staggered magnetic field upon charged particles. The magnetic field arises from the operator $i \left[M, N \right]$, which provides tunneling with imaginary hopping amplitudes 
\beq
\label{IV.3}
i [M,N] \, = \,2 i \sum_{\tb{r} \in \mathbb{A}, \nu=1-4} \lt\{(-1)^{\nu} a_{\tb{r}}^{\dag} b_{\tb{r}+\tb{e}_{\nu}}
- \tr{H.c.} \rt \}\, .
\eeq
Notably, for $K \approx J$ the magnetic flux per plaquette can be on the order of a fundamental flux quantum, a regime accessible in solid state lattices only at the several 100 Tesla level. It has been shown in Ref.~\cite{Lim:08} that the Hamiltonian  $-J T_{+} - i K \frac{1}{2} \left[M, N \right] + H_{\textrm{int}}$ possesses a rich zero temperature phase diagram. Depending on the tunneling parameters $J$ and $K$, four superfluid phases with distinct symmetries of the corresponding order parameters can be formed (see Fig.~\ref{Fig.3}). Unfortunately, since $\xi_M < J$ and $\xi_N/\hbar \O \ll 1$, the staggered vortex superfluid predicted in Fig.~\ref{Fig.3} if $|K|>|J|$ cannot be accessed in the weak driving regime. Furthermore, since $J$ is positive, the entire left half-plane of the phase diagram in Fig.~\ref{Fig.3} cannot be explored. This raises the question whether strong driving permits to access an extended portion of the phase diagram or rather gives rise to entirely new physics.

\begin{figure}
\includegraphics[scale = .50, angle=0, origin=c]{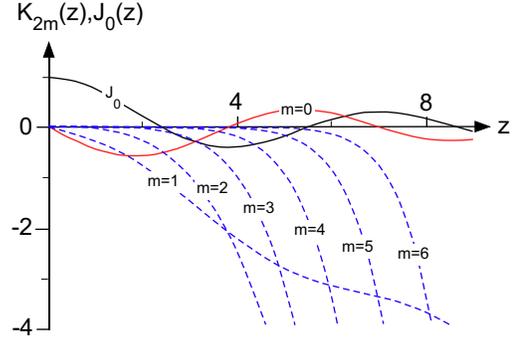}
\caption{\label{Fig.4} (color online) Plot of scaling functions $K_{2m}(z)$ for $m=0,1,2,3,4,5,6$ and $J_0(z)$.} 
\end{figure}

In the following, Eq.~(\ref{II.9}) is used to explore the regime of strong driving for two different choices of the operator $F(t)$. Firstly, the regime of strong driving with respect to $\xi_N N \cos(\O t)$ is explored, i.e. $\xi_N \simeq \hbar \Omega$ while $J,U,\xi_M  \ll \hbar \Omega$. The choice $H_{<}(t) \equiv -J T_{+}  + H_{\tr{int}}  -\xi_M M \sin(\O t)$, $H_{>}(t) \equiv \xi_N N \cos(\O t)$, and $F\equiv  \xi_N N \sin(\O t)/\hbar \Omega$ yields $\hbar F' = H_{>}(t)$, $\left[F,H_{>}(t)\right]=0$ and $\left[F,H_{\tr{int}}\right]=0$. With these settings Eq.~(\ref{II.9}) becomes
\beq
\label{IV.4}
\mathfrak{H}_{F}^{(0,0)} = H_{\tr{int}} &-& J \sum_{n=0}^{\infty} \frac{i^n}{n!}\sin^{n}(\O t) \left(\frac{\xi_N}{\hbar \O}\right)^{n}\left[N, T_{+} \right]_{n}  \nn \\ 
\\ \nn
  &+& \xi_M  \sum_{n=0}^{\infty} \frac{i^n}{n!}\sin^{n+1}(\O t) \left(\frac{\xi_N}{\hbar \O}\right)^{n} \left[N, M  \right]_{n}\,.
\eeq
Using the relations $\left[N,T_{\pm}\right] = -2 T_{\mp}$ and $\left[N, \left[N, M \right] \right] = 4 M$ shown in the appendix, the multiple commutators in Eq.~(\ref{IV.4}) are calculated to be $\left[N,T_{\pm}\right]_n = (-2)^n T_{\pm (-1)^n}$ and $\left[N,M\right]_n =  2^{n-[1-(-1)^n]/2} \left[N,M\right]_{[1-(-1)^n ] /2}\,$. Inserting these into Eq.~(\ref{IV.4}) and making use of Eqs.~(\ref{IIIA.3}) and (\ref{IIIA.4}) leads to
\beq
\label{IV.5}
\langle e^{i2m\Omega t} \mathfrak{H}_{F}^{(0,0)}\rangle_T  = \delta_{m,0} \, H_{\textrm{int}} \qquad\qquad\qquad\qquad \nn  \\  \nn \\ 
- J \, J_{2m}\left(\frac{2\xi_N}{\hbar \Omega}\right) \, T_{+}  - i \frac{1}{2} \xi_M K_{2m}\left(\frac{2\xi_N}{\hbar \Omega}\right) \, \left[N,M\right] \,, \nn  \\  \\  
\textrm{and} \qquad\qquad\qquad\qquad\qquad\qquad\qquad\qquad\qquad\qquad\qquad \nn \\ \nn 
\langle e^{i(2m+1)\Omega t} \mathfrak{H}_{F}^{(0,0)}\rangle_T  =  \qquad\qquad\qquad\qquad\qquad \nn  \\ \nn \\ 
- J \, J_{2m+1}\left(\frac{2\xi_N}{\hbar \Omega}\right) \, T_{-}  + i \xi_M K_{2m+1}\left(\frac{2\xi_N}{\hbar \Omega}\right) \,M \, , \nn 
\eeq 
where $K_{m}(z), m\in \{0,1,...\}$ is defined by
\beq
\label{IV.6}
K_{m}(z)æ\equiv \frac{\partial}{\partial z} \left(J_{m}\left(z\right)-\frac{1+(-1)^m}{2}\frac{z^m}{2^m m!}  \right)   . 
\eeq
In particular, for m=0 one obtains
\beq
\label{IV.7}
H_{\textrm{eff}}æ= &-&J \, J_{0}\left(\frac{2\xi_N}{\hbar \Omega}\right) \, T_{+}  \, \, +  \, \,  H_{\textrm{int}} \, \nn \\
&-& i \frac{1}{2} \xi_M J_{1}\left(\frac{2\xi_N}{\hbar \Omega}\right)  \, \left[M,N\right] \, .
\eeq
Since for $z \ll 1$ (corresponding to weak driving) $J_1(z) \approx z/2 = \xi_N/\hbar \O$, Eq.~(\ref{IV.7}) reproduces the result $H_{\textrm{eff}} \approx -J T_{+} - K \frac{1}{2} \left[M, N \right]$ found in Refs.~\cite{Lim:08, Lim:10}. In the regime of strong driving the formal structure of the effective Hamiltonian is maintained, however with rescaled tunneling energies. The scaling functions $J_0(z)$ and $K_0(z) = - J_1(z)$ illustrated in Fig.~\ref{Fig.4} (solid lines) are bounded with maximal values smaller than unity. They exhibit zero crossings, which permits to suppress $J_{\tr{eff}}\equiv J J_0(z)$ and $K_{\tr{eff}}\equiv K K_0(z)$ or even adjust negative values. 

Upon use of the arguments discussed at the end of subsection~\ref{subsec3a}, equations~(\ref{IV.5}) and (\ref{IV.7}) imply that the description in terms of $H_{\textrm{eff}}$ is suitable if the values of the parameters $J,U,\xi_M$ are constrained by the relations $J J_m(z),U,\xi_M K_m(z) \ll \hbar \O$. Since $ J_m(z) \leq 1$ and $K_{2m+1}(z) \leq 1/2$  for all $m= 0,1,2,...$, it is sufficient to require $J, U, \xi_M, \xi_M K_{2m}(z) \ll \hbar \O$. Allthough $K_{2m}(z)$ is not bounded, one may allow $K_{2m}(z)$ to be significantly larger than unity without affecting the necessary condition $\xi_M K_{2m}(z) \ll \hbar \O$, since $\xi_M \ll \hbar \O$. This in turn permits values of $z$ significantly exceeding unity, i.e., within the regime of strong driving. This is illustrated in Fig.~\ref{Fig.4}, where $K_{2m}(z)$ is plotted for $m=0,1,2,3,4,5,6$.
One recognizes that, in the area around $z \approx 2$, where $|K_0(z)|$ attains its maximum, the values of $K_{2m}(z)$ with $m > 0$ remain well below one, i.e., the requirement to prevent transitions between different Floquet bands is satisfied. In the region $2< z <2.5$ the effective hopping amplitude $J_{\tr{eff}}$ can be tuned close to zero, while $K_0(z)$ maintains sizable values. This is the regime of interest in experiments exploring the staggered vortex superfluid phase of Fig.~\ref{Fig.3}, where tunneling is dominated by the staggered magnetic flux ($K_{\tr{eff}}> J_{\tr{eff}}$).

Although, the rotor potential defined in the context of Eq.~(\ref{IV.1}) constrains $\xi_M$ to be on the order of or smaller than $J$ with the consequence that $\xi_M \ll \hbar \O$, it is nevertheless instructive to examine an alternative approach to an effective Hamiltonian, which does not a priori presuppose small values of $\xi_M$. This amounts to the choice $H_{<}(t) \equiv -J T_{+}$ and $H_{>}(t) \equiv W(t)$. For simplicities sake, the interaction $H_{\tr{int}}$ is neglected here. It turns out practical to rewrite the driving term as 
\beq
\label{IV.8}
W(t) &=& Q^{\dag} e^{i\O t} + Q e^{-i\O t}\,,   \\ 
Q &\equiv&  \frac{1}{2} (\xi_N N+ i \xi_M M)\,. \nn
\eeq
With $F= i(Q^{\dag} e^{i \Omega t} - Q e^{-i \Omega t})/\hbar \Omega$ and thus $\hbar F' = W$ one finds $\left[F, W\right]= 2i \left[ Q, Q^{\dag} \right] / \hbar \Omega$, which inserted into Eq.~(\ref{II.9}) yields
\beq
\label{IV.9}
\mathfrak{H}_{F}^{(0,0)} &=& - J \sum_{n=0}^{\infty} \frac{i^n}{n!} \left[F,T_{+}\right]_{n}  \\ 
&-& \frac{2}{\hbar \Omega } \sum_{n=0}^{\infty} \frac{i^n}{n! (n+2)} \left[F, \left[Q, Q^{\dag} \right] \right]_{n}\, . \nn 
\eeq
Evaluation of $H_{\textrm{eff}}$ requires to calculate the multiple commutators $\left[F,T_{+}\right]_{n}$ and $\left[F, \left[Q, Q^{\dag} \right] \right]_{n}$. 

\textit{Calculation of $\left[F,T_{+}\right]_{n}$}:
a calculation outlined in the appendix yields the commutators $\left[M,T_{\pm}\right] = (1~\mp~1)\, \sum_{\tb{r}\in \mathbb{A}, \nu=1-4} (-1)^{\nu+1} \left(a_{\tb{r}}^{\dag}a_{\tb{r}+2\tb{e}_{\nu}} - b_{\tb{r}-\tb{e}_{\nu}}^{\dag}b_{\tb{r}+\tb{e}_{\nu}}   \right)$. 
While this implies that $\left[M,T_{+}\right] = 0$, one notes that $\left[M,T_{-}\right]$ exclusively comprises terms denoting non-local tunneling between distant lattice sites. Such terms should not significantly contribute and are henceforth neglected, and thus $\left[M,T_{-}\right] \approx 0$ is assumed. In addition, the commutator relation $\left[N,T_{\pm}\right]_n = (-2)^n T_{\pm (-1)^n}$, already applied below Eq.~(\ref{IV.4}), holds. Thus, the summand proportional to $M$ in $F = \left[N \xi_N \sin(\Omega t) - M \xi_M  \cos(\Omega t)\right]/ \hbar \Omega$ can not contribute to $\left[F,T_{+}\right]_{n}$, i.e., 
\beq
\label{IV.10}
\left[F,T_{+}\right]_{n} = \left(\frac{-2 \xi_N}{\hbar\Omega}\right)^{n} \sin^n(\Omega t) \, T_{(-1)^n} \, .
\eeq

\textit{Calculation of $\left[F, \left[Q, Q^{\dag} \right] \right]_{n}$}: since $\left[Q, Q^{\dag} \right] = -\frac{i}{2} \xi_N \xi_M \left[N, M \right]$, multiple commutators of the form $\left[F, \left[N, M \right] \right]_{n}$ have to be considered. A direct calculation (cf. appendix) yields $\left[N, \left[N, M \right] \right] = 4 M$  and $\left[M, \left[M, N \right] \right] = 16 N + O_{\tr{NL}}$, where $O_{\tr{NL}}$ denotes the non-local tunneling terms specified in Eq.~(\ref{app2}), which are henceforth neglected. Introducing the abbreviations $S \equiv \frac{1}{2}\left[N, M \right]$ and $L \equiv \frac{1}{2} M$ yields the relations
\beq
\label{IV.11}
\left[N, L \right]  &=& S \nn \\ 
\left[N, S \right]  &=& 4 L     \\ 
\left[L, S \right]  &=& -4 L \nn 
\eeq
Because $F = p_L L + p_N N$ is linear in $L$ and $N$ with $p_L = -\frac{2 \xi_M}{\hbar \Omega} \cos(\Omega t)$, and $p_N = \frac{\xi_N}{\hbar \Omega}\sin(\Omega t)$, the multiple commutator $\left[F, S \right]_{n}$ is a sum of commutators $\left[F_1, \dots \left[F_n , S \right]\right]$ where $F_{\nu}, \nu \in \{1,\dots,n\}$ denote either of the operators $p_L L$ or $p_N N$. According to the operator algebra of Eq.~(\ref{IV.11}), for even $n$ all possible combinations $\left[F_n, \dots \left[F_1 , S \right]\right]$ are scalar multiples of $S$, while for odd $n$ scalar multiples of $N$ or $L$ can arise. Furthermore, only those contributions $\left[F_n, \dots \left[F_1 , S \right]\right]$ are non-zero, for which, except for $F_n$ in case of odd $n$, the $F_{\nu}$'s occur in pairs, i.e, $F_{2\nu-1} = F_{2\nu}$. Summing over all contributions yields after some algebra
\beq
\label{IV.12}
\left[F, \left[Q, Q^{\dag} \right] \right]_{2n} &=& -i \xi_N \xi_M 4^n \left(p_L^2 + p_N^2 \right)^n S \nn \\ 
\left[F, \left[Q, Q^{\dag} \right] \right]_{2n+1} &=& \\ \nn
-i \xi_N \xi_M &4^{n+1}& \left(p_L^2 + p_N^2 \right)^m \left(  p_N  L -p_L  N \right)  \, ,
\eeq
where $n \in \{0,1, ... \}$. 

\begin{figure}
\includegraphics[scale = .40, angle=0, origin=c]{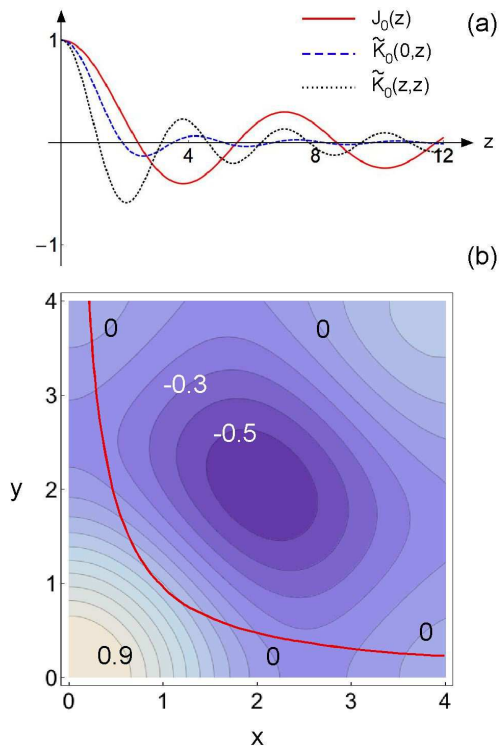}
\caption{\label{Fig.5} (color online) (a) Plot of scaling functions $J_0(z)$ (red, solid), $\tilde{K}_0(z,0)$ (blue, dashed), and $\tilde{K}_0(z,z)$ (black, dotted). (b) Plot of scaling function $\tilde{K}_0(x,y)$. The thick red line is given by $x y = 0.2$.} 
\end{figure}

With the expressions of Eq.~(\ref{IV.10}) and Eq.~(\ref{IV.12}) we are now in the position to calculate $H_{\textrm{eff}}$ by means of Eq.~(\ref{IV.9}). Upon noting that the time-averages of the odd commutators yield zero contributions, i.e., $\langle \left[F, T_{+} \right]_{2n+1}\rangle_T =  \langle \left[F, \left[Q, Q^{\dag} \right] \right]_{2n+1}\rangle_T = 0$, and that for arbitrary $x,y \in \mathbb{R}$
\beq
\label{IV.13}
\langle \left( x^2 \sin^2(\Omega t)+y^2 \cos^2(\Omega t)\right)^n \rangle_{T} = \qquad\qquad\qquad\qquad \\ \nn
 \sum_{\nu=0}^n \left(2(n-\nu) \atop n-\nu \right)\left(2\nu \atop \nu \right) \left(\frac{x}{2}\right)^{2(n-\nu)} \left(\frac{y}{2}\right)^{2\nu}\, ,
\eeq
one obtains 
\beq
\label{IV.14}
H_{\textrm{eff}} &=& - \tilde{J}_{\tr{eff}}\, T_{+} -  i \,\tilde{K}_{\tr{eff}}\, \frac{1}{2}\, \left[M, N \right]  \nn \\ 
\tilde{J}_{\tr{eff}}&\equiv& J \, J_0\left(\frac{2 \xi_N}{\hbar \Omega} \right) \\ \nn
\tilde{K}_{\tr{eff}} &\equiv& K\, \tilde{K}_0\left(\frac{2 \xi_M}{\hbar \Omega}, \frac{\xi_N}{\hbar \Omega} \right) \\ \nn \, .
\eeq
Here, $J_0(z)$ denotes the Bessel function of zero order and
\beq
\label{IV.15}
\tilde{K}_0(x,y) \, \, \equiv \, \, \sum_{n=0}^{\infty}\sum_{\nu=0}^{n}\,\,\frac{(-4)^n}{(n+1)(2n)!} \, \, \times      \qquad\qquad\qquad   \nn \\ \\
\left(2(n-\nu) \atop n-\nu \right)\left(2\nu \atop \nu \right) \left(\frac{x}{2}\right)^{2(n-\nu)}\left(\frac{y}{2}\right)^{2\nu}
 \, .\nn 
\eeq
Note the symmetry relation $\tilde{K}_0(x,y) = \tilde{K}_0(y,x)$. Furthermore, $\tilde{K}_0(x,x) = \sin(2 x)/x - \sin^2(x)/x^2$ and $\tilde{K}_0(0,y) = J_1(2 y)/y$. As a consequence of the latter equation,  if  $\tilde{K}_0$ is constrained to the $y$-axis, one recovers the case of small $\xi_M \ll \hbar \Omega$ described by Eq.~(\ref{IV.7}). The scaling functions $J_0(z)$ and $K_0(x,y)$ illustrated in Fig.~\ref{Fig.5} are bounded with maximal values of unity arising at the origin. In contrast to $\tilde{J}_{\tr{eff}}$, $\tilde{K}_{\tr{eff}}$ is not bounded with respect to the driving parameters $\xi_M$ and $\xi_N$ since $K$ itself scales with $\xi_M \xi_N$. In order to determine the range of $\xi_M$ and $\xi_N$ for which conditions~(\ref{II.4}) and (\ref{II.6}) hold, $\langle e^{i(m'-m)\Omega t} \mathfrak{H}_{F}^{(0,0)}\rangle_T$ can be readily calculated by means of Eq.~(\ref{IV.12}) and variants of Eq.~(\ref{IV.13}). With the same relaxed variants of conditions~(\ref{II.4}) and (\ref{II.6}) used in the previous discussion one finds that, either of the parameters $\xi_M/\hbar\Omega$ and $\xi_N/\hbar\Omega$ may exceed unity, as long as their product remains well below unity. To illustrate the area of permissible values for $\xi_M$ and $\xi_N$, the red line in Fig.~\ref{Fig.5}(b) shows the upper boundary of the region defined by $\xi_M \xi_N \leq 0.1 (\hbar\Omega)^2$. When passing too far into the region above this line, one may encounter transitions between different Floquet bands such that the description in terms of the effective Hamiltonian $H_{\textrm{eff}}$ fails. Recall, that at two points in the calculations leading to Eq.~(\ref{IV.14}) non-local terms have been neglected, which describe tunneling between distant lattice sites. These terms arise in quadratic or higher order with respect to $\xi_M/ \hbar \O$, such that in the calculation of Eq.~(\ref{IV.7}) they were a priori excluded by the assumption $\xi_M/ \hbar \O \ll 1$. Finally, if the Hubbard interaction $H_{\textrm{int}}$ had been accounted for in the derivation of Eq.~(\ref{IV.14}), the choice $\hbar F'=W$ would have led to a non-zero commutator $[F,H_{\textrm{int}}]$ with the consequence of non-local interaction terms also scaling with $\xi_M/ \hbar \O$ to second or higher order.     

\section{Conclusions} \label{conc}
In this article, a general expression for the effective Hamiltonian of a periodically driven quantum system has been derived and applied to various optical lattice models with external driving, which are readily accessible in experiments. Three scenarios have been considered: a periodically shifted 1D lattice, a 2D square lattice with anticyclic temporal modulation of the well depths of adjacent lattice sites, and a 2D lattice, subjected to an array of microscopic rotations commensurate with its plaquette structure. For the 1D scenario, the rescaling of the tunneling strength previously found by Eckardt et al. in Ref.~\cite{Eck:05} is reproduced, showing that strong driving permits suppression of tunneling and the adjustment of negative tunneling energies. Well depth modulation inversely phased for adjacent lattice sites lets us extend these effects to a 2D scenario. In the case of the 2D square lattice with staggered rotation, the expression previously found for weak driving by Lim et al. in Refs.~\cite{Lim:08,Lim:10} is generalized. The effective Hamiltonian is found to maintain its formal structure independent of the driving strength, i.e., the interpretation that it simulates the effect of a staggered magnetic field upon charged particles, previously found for weak driving, extends to the regime of strong driving. For all lattice scenarios one finds that the regime of strong driving differs from that of weak driving essentially by an energy rescaling. In case of the 2D lattice with staggered rotation, this rescaling should permit to extend the experimentally accessible portion of the phase diagram. The general expression for the effective Hamiltonian might also be useful in numerous cases of interest not considered here.

\section*{Acknowledgements} 
It is a pleasure to acknowledge several very helpful discussions with Andr\'{e} Eckardt. He has drawn my attention to the specific adequacy of the small $\xi_M$ approximation underlying Eq.~(\ref{IV.4}). I also thank Cristiane Morais Smith for useful comments and for proofreading the manuscript. This work was partially supported by DFG-grant He2334/10-1, DFG GrK1355, and Joachim Herz Stiftung Hamburg.

\appendix
\section{}\label{App}
The following calculations are carried out for bosons, i.e., $\left[a_{\tb{r}},a_{\tb{r'}}^{\dag}\right] = \left[b_{\tb{r}},b_{\tb{r'}}^{\dag}\right] = \delta_{\tb{r},\tb{r'}}$, while all other possible commutators are zero. Similar calculations yield analog results for fermions. The following definitions are used
\beq
\label{app1}
&R& \equiv \sum_{\tb{r}\in \mathbb{A} }\sum_{\nu\in {1,2,3,4}} a_{\tb{r}} b_{\tb{r}+\tb{e}_{\nu}}^{\dag} \\ \nn 
&S& \equiv \sum_{\tb{r}\in \mathbb{A} }\sum_{\nu\in {1,2,3,4}} (-1)^{\nu+1} a_{\tb{r}} b_{\tb{r}+\tb{e}_{\nu}}^{\dag} \\ \nn
&N_{A}&\equiv \sum_{\tb{r}\in \mathbb{A} } a_{\tb{r}}^{\dag} a_{\tb{r}} \,\,, N_{B} \equiv \sum_{\tb{r}\in \mathbb{B} } b_{\tb{r}}^{\dag} b_{\tb{r}} \\ \nn 
 &T_{\pm}& \equiv R \pm R^{\dag}\, , M \equiv S + S^{\dag}, N \equiv N_A -N_B\,.
\eeq

\textit{Calculation of $\left[M,T_{\pm}\right]$}: with the definitions of Eq.~(\ref{app1}) and $\left[R,S \right] = \left[R^{\dag},S^{\dag} \right] = 0$ it follows $\left[M,T_{\pm} \right] = \left[S^{\dag},R\right] \mp \left[S^{\dag},R\right]^{\dag}$. One obtains five kinds of contributions to $\left[S^{\dag},R\right]$: 1) hopping terms connecting nearest neighbor $\mathbb{A}$-sites $\left[ (-1)^{\nu+1} a_{\tb{r}}^{\dag} b_{\tb{r}+\tb{e}_{\nu}}, a_{\tb{r}+\tb{e}_{\nu}+\tb{e}_{\mu}} b_{\tb{r}+\tb{e}_{\nu}}^{\dag}  \right] = (-1)^{\nu+1} a_{\tb{r}}^{\dag} a_{\tb{r}+\tb{e}_{\nu}+\tb{e}_{\mu}}$ with odd $\nu+\mu \,$; 2) hopping terms connecting nearest neighbor $\mathbb{B}$-sites $\left[ (-1)^{\nu+1} a_{\tb{r}}^{\dag} b_{\tb{r}+\tb{e}_{\nu}}, a_{\tb{r}} b_{\tb{r}+\tb{e}_{\mu}}^{\dag} \right] = -(-1)^{\nu+1} b_{\tb{r}+\tb{e}_{\mu}}^{\dag} b_{\tb{r}+\tb{e}_{\nu}}$ with odd $\nu+\mu \,$; 3) hopping terms connecting next nearest neighbor $\mathbb{A}$-sites $\left[ (-1)^{\nu+1} a_{\tb{r}}^{\dag} b_{\tb{r}+\tb{e}_{\nu}}, a_{\tb{r}+2\tb{e}_{\nu}} b_{\tb{r}+\tb{e}_{\nu}}^{\dag}  \right] = (-1)^{\nu+1} a_{\tb{r}}^{\dag} a_{\tb{r}+2\tb{e}_{\nu}} \,$; 4) hopping terms connecting next nearest neighbor $\mathbb{B}$-sites $\left[ (-1)^{\nu+1} a_{\tb{r}}^{\dag} b_{\tb{r}+\tb{e}_{\nu}}, a_{\tb{r}} b_{\tb{r}-\tb{e}_{\nu}}^{\dag} \right] = (-1)^{\nu+1} b_{\tb{r}-\tb{e}_{\nu}}^{\dag} b_{\tb{r}+\tb{e}_{\nu}} \,$; 5) on-site terms $\left[ (-1)^{\nu+1} a_{\tb{r}}^{\dag} b_{\tb{r}+\tb{e}_{\nu}}, a_{\tb{r}} b_{\tb{r}+\tb{e}_{\nu}}^{\dag} \right] = (-1)^{\nu+1} \left(a_{\tb{r}}^{\dag}a_{\tb{r}} - b_{\tb{r}+\tb{e}_{\nu}}^{\dag} b_{\tb{r}+\tb{e}_{\nu}} \right)$. Only the terms of type 3) or 4) can yield non-zero contributions, if the sum $\tb{r}\in \mathbb{A}$ and $\nu\in {1,2,3,4}$ is carried out, leading to $\left[S^{\dag},R\right] = \sum_{\tb{r}\in \mathbb{A},\nu={1-4}} (-1)^{\nu+1} \left( a_{\tb{r}}^{\dag} a_{\tb{r}+2\tb{e}_{\nu}} - b_{\tb{r}}^{\dag} b_{\tb{r}+2\tb{e}_{\nu}}\right)$. Observing that $\left[S^{\dag},R\right]^{\dag} = \left[S^{\dag},R\right]$, one finally obtains $\left[M,T_{\pm} \right] =\left(1\mp1\right) \sum_{\tb{r}\in \mathbb{A},\nu = {1-4}}(-1)^{\nu+1} \left( a_{\tb{r}}^{\dag} a_{\tb{r}+2\tb{e}_{\nu}} - b_{\tb{r}}^{\dag} b_{\tb{r}+2\tb{e}_{\nu}}\right)$.

\textit{Calculation of $\left[N,T_{\pm}\right]$}: With the definitions of Eq.~(\ref{app1}) one calculates $\left[N_A,T_{\pm}\right] = 
\left[N_A,R\right]\mp\left[N_A,R\right]^{\dag}$. With $\left[N_A,R\right] = \sum_{\tb{r}\in \mathbb{A},\nu = {1-4}} \left[a_{\tb{r}}^{\dag} a_{\tb{r}}, a_{\tb{r}} b_{\tb{r}+\tb{e}_{\nu}}^{\dag}\right] = - R$ one obtains $\left[N_A,T_{\pm}\right] =-T_{\mp}$. A similar calculation gives $\left[N_B,T_{\pm}\right] = T_{\pm}$ and thus $\left[N,T_{\pm}\right] =-2T_{\mp}$.

\textit{Calculation of $\left[N,\left[M,N\right]\right]$}:
Using the definitions of Eq.~(\ref{app1}) one calculates $\left[M,N\right] = 2\left(S-S^{\dag}\right)$. With $\left[N_A,\left[M,N\right]\right] = 2 \left( \left[N_A,S\right] +\left[N_A,S\right]^{\dag}\right)$ and $\left[N_A,S\right] = \sum_{\tb{r}\in \mathbb{A},\nu = {1-4}}  (-1)^{\nu+1} \left[a_{\tb{r}}^{\dag} a_{\tb{r}}, a_{\tb{r}} b_{\tb{r}+\tb{e}_{\nu}}^{\dag}\right] = - S$ one obtains $\left[N_A,\left[M,N\right]\right] = - 2M$. An analog calculation yields $\left[N_B,\left[M,N\right]\right] = 2M$ and thus $\left[N,\left[M,N\right]\right] = - 4M$.

\textit{Calculation of $\left[M,\left[M,N\right]\right]$}: With $\left[M,N\right] = 2\left(S-S^{\dag}\right)$ one gets $\left[M,\left[M,N\right]\right] = 4 \left[S^{\dag},S\right]$. Five kinds of contributions to $\left[S^{\dag},S\right]$ arise: 1) hopping terms connecting nearest neighbor $\mathbb{A}$-sites $\left[ (-1)^{\nu+1} a_{\tb{r}}^{\dag} b_{\tb{r}+\tb{e}_{\nu}}, (-1)^{\mu+1} a_{\tb{r}+\tb{e}_{\nu}+\tb{e}_{\mu}} b_{\tb{r}+\tb{e}_{\nu}}^{\dag}  \right] = - a_{\tb{r}}^{\dag} a_{\tb{r}+\tb{e}_{\nu}+\tb{e}_{\mu}}$ with odd $\nu+\mu \,$; 2) hopping terms connecting nearest neighbor $\mathbb{B}$-sites $\left[ (-1)^{\nu+1} a_{\tb{r}}^{\dag} b_{\tb{r}+\tb{e}_{\nu}}, (-1)^{\mu+1} a_{\tb{r}} b_{\tb{r}+\tb{e}_{\mu}}^{\dag} \right] = b_{\tb{r}+\tb{e}_{\mu}}^{\dag} b_{\tb{r}+\tb{e}_{\nu}}$ with odd $\nu+\mu \,$; 3) hopping terms connecting next nearest neighbor $\mathbb{A}$-sites $\left[ (-1)^{\nu+1} a_{\tb{r}}^{\dag} b_{\tb{r}+\tb{e}_{\nu}}, (-1)^{\nu+1} a_{\tb{r}+2\tb{e}_{\nu}} b_{\tb{r}+\tb{e}_{\nu}}^{\dag}  \right] = a_{\tb{r}}^{\dag} a_{\tb{r}+2\tb{e}_{\nu}} \,$; 4) hopping terms connecting next nearest neighbor $\mathbb{B}$-sites $\left[ (-1)^{\nu+1} a_{\tb{r}}^{\dag} b_{\tb{r}+\tb{e}_{\nu}}, (-1)^{\nu+1} a_{\tb{r}} b_{\tb{r}-\tb{e}_{\nu}}^{\dag} \right] = - b_{\tb{r}-\tb{e}_{\nu}}^{\dag} b_{\tb{r}+\tb{e}_{\nu}} \,$; 5) on-site terms $\left[ (-1)^{\nu+1} a_{\tb{r}}^{\dag} b_{\tb{r}+\tb{e}_{\nu}}, (-1)^{\nu+1} a_{\tb{r}} b_{\tb{r}+\tb{e}_{\nu}}^{\dag} \right] = a_{\tb{r}}^{\dag}a_{\tb{r}} - b_{\tb{r}+\tb{e}_{\nu}}^{\dag} b_{\tb{r}+\tb{e}_{\nu}}\,$. Suming up over all $\tb{r}\in \mathbb{A}$ and $\nu,\mu \in {1,2,3,4}$ in compliance with the constraint of odd $\nu+\mu \,$ for type 1) and 2) terms yields
\beq
\label{app2}
\left[M,\left[M,N\right]\right] = \qquad\qquad\qquad\qquad\qquad\qquad\qquad  \\ \nn
16 N \,\,+\, \,4 \sum_{\tb{r}\in \mathbb{A} \atop {\nu,\mu \in {1,2,3,4} \atop |\nu - \mu| \neq 2}} a_{\tb{r}}^{\dag}a_{\tb{r}+\tb{e}_{\nu}+\tb{e}_{\mu}}-b_{\tb{r}}^{\dag}b_{\tb{r}+\tb{e}_{\nu}+\tb{e}_{\mu}}\, .
\eeq
The first term $16 N$ results from the summation of the on-site contributions in 5), while all other contributions of type 1)-4) are collected in the sum of non-local tunneling terms.

\end{document}